\documentclass[review]{elsarticle}

\usepackage[left=3cm, right=3cm, top=3cm]{geometry}
\usepackage{lineno,hyperref,amsmath,amsthm,esvect}

\modulolinenumbers[5]

\journal{International Journal of Computer Mathematics: Computer Systems Theory}

\newtheorem{corollary}{Corollary}[section]
\newtheorem{theorem}{Theorem}[section]
\newtheorem{lemma}{Lemma}[section]

\graphicspath{ {Figures/} }

\begin{document}

\begin{frontmatter}

\title{\bf\large On approximations to minimum link visibility paths in simple polygons}

\author[mymainaddress]{Mohammad Reza Zarrabi}
\ead{m.zarabi@modares.ac.ir}

\author[mymainaddress]{Nasrollah Moghaddam Charkari\corref{mycorrespondingauthor}}
\cortext[mycorrespondingauthor]{Corresponding author}
\ead{charkari@modares.ac.ir}

\address[mymainaddress]{Faculty of Electrical Engineering and Computer Science, Tarbiat Modares University, Tehran, Iran}

\begin{abstract}
We investigate a practical variant of the well-known polygonal visibility path (watchman) problem.
For a polygon $P$, a minimum link visibility path is a polygonal visibility path in $P$
that has the minimum number of links.
The problem of finding a minimum link visibility path is NP-hard for simple polygons.
If the link-length (number of links) of a minimum link visibility path (tour) is $Opt$
for a simple polygon $P$ with $n$ vertices, we provide an algorithm with $O(kn^2)$ runtime that
produces polygonal visibility paths (or tours) of
link-length at most $(\gamma+a_l/(k-1))Opt$ (or $(\gamma+a_l/k)Opt$), where $k$ is a parameter dependent on $P$,
$a_l$ is an output sensitive parameter and
$\gamma$ is the approximation factor of an $O(k^3)$ time approximation algorithm for the
graphic traveling salesman problem (path or tour version).
\end{abstract}


\end{frontmatter}

\section{Introduction}
The \emph{polygonal visibility path (watchman)} problem is a frequently studied topic in
computational geometry and optimization.
This problem is motivated by many applications such as security and surveillance (e.g., guarding, exploring and analyzing buildings and areas), saving energy and time (e.g., photographing on area with less frames), efficient simulations and more.
Two points in a polygon $P$ are \emph{visible} to each other if their connecting segment remains completely inside $P$.
A polygonal visibility path for a given $P$ is a polygonal path contained in $P$ with the
property that every point inside $P$ is visible from at least one point on the path.
A \emph{minimum link visibility path}
is a polygonal visibility path in $P$ with the minimum \emph{link-length} (number of \emph{links}).
Minimum link paths appear to be of great importance in robotics and communications
systems, where straight line motion or communication is relatively inexpensive but
turns are costly.

We consider the problem of finding minimum link visibility paths in simple polygons.
Our objective is to compute approximate minimum link visibility paths (tours) in a simple polygon $P$ with $n$ vertices.

This problem has been extensively studied in \cite {Alsuwaiyel_1993,Alsuwaiyel_1995}.
In \cite {Alsuwaiyel_1993} Alsuwaiyel and Lee
showed that the problem is NP-hard by constructing the complex gadgets on the outer boundary of $P$.
They also presented an $O(n^3 \log n)$ time approximation algorithm with constant approximation factors 3 and 2.5
by employing some famous approximation algorithms for the \emph{graphic traveling salesman problem} (GTSP).
However, the approximation algorithm in \cite {Alsuwaiyel_1993} gives a feasible solution with no bound guarantee.
For this reason, another approximation algorithm was provided in \cite {Alsuwaiyel_1995}.
The time complexity of this new approximation algorithm was $O(n^2)$ and $O(n^3)$
with constant approximation factors 4 and 3.5,
respectively.
For a polygonal domain, even with a very simple outer boundary (possibly none), Arkin et al.
showed the problem is NP-complete \cite {Arkin_2003}.
Also, they gave a polynomial time approximation algorithm with the
approximation factor $O(\log n)$, where $n$ is the number of vertices in the polygonal domain.
A summery of the approximation algorithms are given in Table 1.

\begin{table}[t]
\begin{center}
\small
\begin{tabular}{|ccccccccccc|}
\hline
\emph{Scene} && \emph{Running time} && \emph{Approximation factor} && \emph{Version} && \emph{Validity} && \emph{Ref}\\
\hline
\hline
Polygonal domain && Polynomial && $O(\log n)$ && Tour && True && \cite {Arkin_2003}\\
\hline
Simple polygon && $O(n^3 \log n)$ && 3 && Path && False && \cite {Alsuwaiyel_1993}\\
\hline
Simple polygon && $O(n^3 \log n)$ && 2.5 && Path && False && \cite {Alsuwaiyel_1993}\\
\hline
Simple polygon && $O(n^2)$ && 4 && Path/Tour && True && \cite {Alsuwaiyel_1995}\\
\hline
Simple polygon && $O(n^3)$ && 3.5 && Path/Tour && True && \cite {Alsuwaiyel_1995}\\
\hline
\end{tabular}
\end{center}
\caption{A summary of the approximation algorithms
for finding a minimum link visibility path in polygons.}
\end{table}

In this paper, we modify the approximation algorithm presented by Alsuwaiyel and Lee \cite {Alsuwaiyel_1995}, using the concepts introduced by Zarrabi and Charkari \cite {Zarrabi_2020_1}.
Briefly, we compute geometric loci of points (called a \emph{Cell})
whose sum of link distances to the source and destination is constant.
This is done by the \emph{shortest path map} and \emph{map overlay} techniques \cite {Suri_1990, Edelsbrunner_1986}.
The minimum of these sums is then considered for each iteration in the modified algorithm
with the time complexity $O(kn^2)$ ($k$ is at most the number of nonredundant cuts of $P$).
As a result, polygonal visibility paths (or tours) of link-length at most
$(1.5+a_l/(k-1))Opt$ (or $(1.4+a_l/k)Opt$) are produced, where $a_l$ is an output sensitive parameter
and $Opt$ is the link-length of a minimum link visibility path in $P$.
Note that $a_l/(k-1)$ (or $a_l/k$) is a positive rational number no more than 2, but in most cases it is
less than 2.

\section{Preliminaries}
Let $P$ be a \emph{non-star-shaped} simple polygon
with $n$ vertices, sorted in \emph{clockwise} order.
As mentioned in \cite {Alsuwaiyel_1995} and \cite {Tan_2007}, assume that no three
vertices of $P$ are \emph{collinear} and the extensions of two
non-adjacent edges of $P$ do not intersect at a boundary point, respectively.

With these properties, $P$ is called a \emph{watchman polygon}.
We borrow the related terminology from \cite {Alsuwaiyel_1995, Zarrabi_2020_1, Zarrabi_2020_2}
and review some terms adapted to the notation used in this paper.

\begin{figure}
	\centering
	\includegraphics[width=12.5cm,keepaspectratio=true]{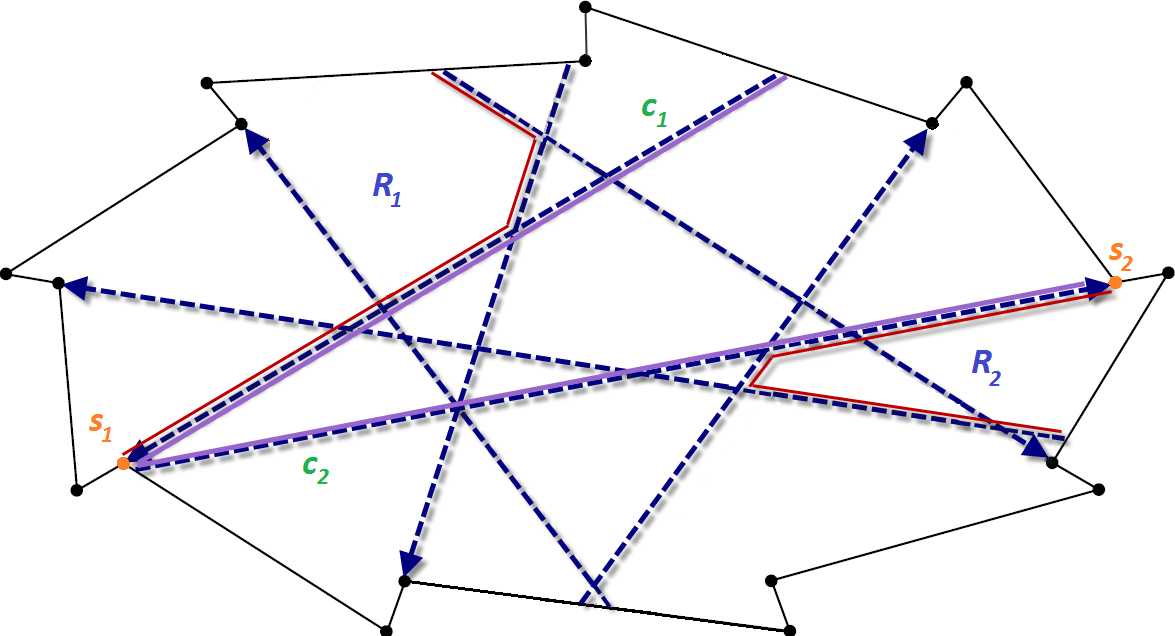}
	\caption{The set $\mathcal{C}$ (nonredundant cuts) consists of all dashed arrows and $M=\{c_1, c_2\}$ (maximum independent set)}
	\label{fig:IndependentCuts}
\end{figure}

Let $v$ be a reflex vertex, $u$ a vertex adjacent to $v$, and $w$ the closest point to $v$
on the boundary of $P$ hit by the half line originating at $v$ along $\overline{uv}$.
Then, we call the line segment $c=\overline{vw}$ a \emph{cut} of $P$.
Since $c$ partitions $P$ into two portions, we define $Pocket(c)$ to be the portion of $P$ that includes $u$.
We also associate a direction to each cut $c$ such that
$Pocket(c)$ entirely lies to the right of $c$ (right hand rule). This direction is compatible with the
clockwise ordering of vertices of $P$ \cite {Alsuwaiyel_1995}.
Start and end points of a directed cut $c$ will be denoted by $\alpha(c)$ and $\beta(c)$, respectively.
If $c$ and $c'$ are two cuts such that $Pocket(c) \subset Pocket(c')$,
then $c'$ is called \emph{redundant}, otherwise it is \emph{nonredundant}.
Let $\mathcal{C}$ be the set of all nonredundant cuts of $P$ (see Figure~\ref{fig:IndependentCuts}).

The cuts in $C$, where $C \subseteq \mathcal{C}$ are called \emph{independent},
if $\forall$ $c,c' \in C$ $(c \neq c')$, then $Pocket(c) \cap Pocket(c') = \emptyset$.
$C$ is a \emph{maximal independent set of cuts},
if $\forall$ $c \in (\mathcal{C} - C)$, then $\exists$ $c' \in C$
such that $c$ and $c'$ are not independent.
A set $M \subseteq \mathcal{C}$ is said to be a \emph{maximum independent set of cuts}, if it is a maximal
independent set of cuts with maximum cardinality
(see Figure~\ref{fig:IndependentCuts}).
The sets $\mathcal{C}$ and $M$ can be computed\footnote
{The endpoints of cuts in $\mathcal{C}$ must be sorted in
\emph{counterclockwise} order in the greedy algorithm for computing $M$.}
in $O(n)$ time according to \cite {Tan_2007} and \cite {Alsuwaiyel_1995}, respectively.

A \emph{visibility path} for a given $P$ is
a connected path (possibly curved) contained in $P$ with the property that
every point inside $P$ is visible from at least one point on the path.
The following theorem is proved in \cite {Zarrabi_2020_2}:

\begin{theorem}\label{Based}
A curved path inside $P$ is a visibility path, iff it intersects all $c \in \mathcal{C}$ (all $Pocket(c)$).
\end{theorem}

A \emph{minimum link} visibility path $\Pi$ is a \emph{polygonal} visibility path in $P$
with the minimum \emph{link-length}.
Thus, a minimum link visibility path must intersect all members of $\mathcal{C}$. Conversely,
if a minimum link path intersects all members of $\mathcal{C}$, then it is a visibility path.
Consequently, it is necessary that $\Pi$ intersects all members of $M$.
Without loss of generality, suppose that $(\mathcal{C}-M) \neq \emptyset$.
Let $n(X)$ be the size of a set $X$.
Note that when all cuts in $\mathcal{C}$ intersect (checkable in $O((n(\mathcal{C}))^2)=O(n^2)$),
$M$ consists of exactly one cut,
i.e., $\Pi$ can be any member of $\mathcal{C}$ with link distance one.
Otherwise, $n(M) > 1$.
Since at least one line segment is needed in order
for $\Pi$ to go from one cut in $M$ to the next \cite {Alsuwaiyel_1995},
$|\Pi| \geq n(M)-1$ ($|.|$ denotes the link-length).
An important property of $M$ is summarized as follows \cite {Alsuwaiyel_1995}:

\begin{lemma}\label{Group}
Suppose that $M = \{c_1, c_2, ..., c_k\}$.
Then, for any $c \in (\mathcal{C}-M)$ there is a cut $c' \in M$
such that $\alpha(c)$ lies in the open interval defined by
$(\alpha(c'),\beta(c'))$.
\end{lemma}

For $c_i \in M$ $(1 \leq i \leq k=n(M))$, $s_i=\beta(c_i)$ is defined. Let $S = \{s_1, s_2, ..., s_k\}$.
The members of $S$ are called \emph{special points}.
According to Lemma~\ref{Group}, $(\mathcal{C}-M)$ partitions into $k$ groups ($\{G_1, G_2, ..., G_k\}$) such that
the special point associated with each $c_i$ lies to the right of the members of each group $G_i$.
Thus, we have the following theorem \cite {Alsuwaiyel_1995}:

\begin{theorem}\label{Special Points}
If a curved path inside $P$ visits all the special points, then it is a visibility path.
\end{theorem}

It is straightforward to see that we can generalize the special points in Theorem~\ref{Special Points} to the
\emph{special regions} $R_i$ (simple polygon),
where every point of $R_i$ lies to the right of the members of each group $G_i$
$(1 \leq i \leq k)$.
Let $R=\{R_1, R_2, ..., R_k\}$ be the set of special regions and let $\partial P$ denote the boundary of $P$.
Each $\partial R_i$ consists of two parts:
one on $\partial P$ and the other inside $P$, which is also a convex chain.
For example, Figure~\ref{fig:IndependentCuts} depicts two sets
$S=\{s_1, s_2\}$ and $R=\{R_1, R_2\}$, where $R_1$ and $R_2$ are bounded by the red convex chains and $\partial P$.
The following corollary is a direct generalization of Theorem~\ref{Special Points}:

\begin{corollary}\label{Special Regions}
If a curved path inside $P$ intersects all the special regions, then it is a visibility path.
\end{corollary}

The convex chain of $\partial R_i$ can be computed in $O((n(G_i))^2)$ time by the simple line intersection algorithm.
On the other hand, since $R_i \subset Pocket(c_i)$ for $c_i \in M$, we have
$\partial R_i \cap \partial R_j = \emptyset$ $(1 \leq i,j \leq k)$.
As a result, $R$ is computed in $O((n(\mathcal{C}))^2)+O(n)=O(n^2)$ time and the total number of
vertices of members of $R$ is $O(n)$ (each $R_i$ has $O(n_i)$ vertices,
where $n_i$ is the number of vertices of $Pocket(c_i)$ and $\sum_{i=1}^{k} n_i=O(n)$).
Unfortunately, Corollary~\ref{Special Regions} is not a necessary condition for $\Pi$.
For this reason, we briefly describe some notation used in \cite {Zarrabi_2020_1} and apply them on the members of $R$.

The notion of \emph{shortest path map} (called SPM) or \emph{window partition} introduced in \cite {Suri_1990} is central to our discussion.
$SPM(x)$ denotes the simply connected planar subdivision of $P$ into \emph{faces}
with the same link distance to a point or a line segment $x$.
Also, $SPM(x)$ has an associated set of \emph{windows}, which are chords of $P$ that serve as boundaries between
adjacent faces.

The construction of $SPM(x)$ takes $O(n)$ time \cite {Suri_1990}.
Consider the two maps $SPM(c_i)$ and $SPM(c_j)$ for cuts $c_i, c_j \in M$.
The link distances from $c_i$ and $c_j$ to the faces of $SPM(c_i)$ and $SPM(c_j)$, respectively
are added to the corresponding faces during their construction.
To compute geometric loci of points whose sum of link distances to $c_i$ and $c_j$ is constant,
the \emph{map overlay} technique is employed, i.e., the intersection of these maps is computed.
As a result, the new simply connected planar subdivision of $P$ is created with new faces.
These faces are called \emph{Cells}.
Construction of \emph{Cells} and computation of their values (defined in Section 3 of \cite {Zarrabi_2020_1})
for arbitrary $c_i$ and $c_j$ are performed in $O(n)$ time
\cite {Zarrabi_2020_1}.
Let $c_h \in M$, where $h \neq i,j$.
Since $c_h$, $c_i$ and $c_j$ are independent, $c_i$ and $c_j$ lie on the same side of $c_h$ inside $P-Pocket(c_h)$
($s$ and $t$ are replaced by $c_i$ and $c_j$, and $p=Pocket(c_h)$ similar to the case $Q(c)$ in \cite {Zarrabi_2020_1}).
Consider the \emph{Cells} or portions of them
lying in $Pocket(c_h)$ for $c_i$ and $c_j$.
Let the minimum value of these \emph{Cells} be denoted by $cellmin_{h,i,j}$, and $Cellmin_{h,i,j}$ be the
set of them with the value $cellmin_{h,i,j}$.
It is proved that $Cellmin_{h,i,j}$ and $cellmin_{h,i,j}$ for the given $c_h$, $c_i$ and $c_j$
are computed in $O(\log n)$ time, after $O(n)$ preprocessing time.
Also, $1 \leq n(Cellmin_{h,i,j}) \leq 2$.
Let $F_{h,i}$ be the face of $SPM(c_i)$ intersecting $c_h$ with minimum value
(each cut intersects at most three faces) and $f_{h,i}=F_{h,i} \cap Pocket(c_h)$.
The number of vertices of $f_{h,i}$ is $O(n_h)$.
Since each line segment intersects at most two windows,
the number of vertices of the member(s) of $Cellmin_{h,i,j}$
would also be $O(n_h)$.
Depending on the positions of $F_{h,i}$, $F_{h,j}$, $\alpha(c_h)$ and $\beta(c_h)$,
one of the six cases $C(a),C(b),C(c_1),C(c_2),C(c_{3_1})$ and $C(c_{3_2})$ may occur \cite {Zarrabi_2020_1}.
We will use these cases in the next section.

\section{Modifications}
First, let us briefly recall the idea of Alsuwaiyel and Lee's algorithm \cite {Alsuwaiyel_1995}.
The heart of this heuristic is based on a well-known algorithm to find an approximate solution of a given instance of the
\emph{traveling salesman problem} and on the property of nonredundant cuts proved in Theorem~\ref{Based}.
Thus, the first step is to compute the set of all nonredundant cuts of $P$.
In the second step, a maximum independent set of these cuts and also
their associated set of special points are computed.
It is important to use such a computation so as to ensure that Lemma~\ref{Group} holds.
In the third step, minimum link path and link distance between each pair of cuts in the maximum independent set
are computed. Then, a complete graph whose node set is the maximum independent set is constructed and
to each edge of this graph the link distance between the corresponding cuts is assigned.
In the fourth step, the Christofides' heuristic \cite {Christofides_1976, Hoogeveen_1991}
is applied to the graph and the corresponding
sequence of cuts are generated as the output of the heuristic.
Finally, minimum link paths between the generated cuts and the intersections between these paths and those cuts,
based on the order of the sequence are considered as a portion of approximate minimum link visibility path.
The desired path is obtained by inserting two additional line segments between the special point
and two intersection points on each of the generated cuts.

We modify mainly the last step of this algorithm based on the approximation algorithms for the GTSP (path or tour version)
whose time complexities are $O(k^3)$, where $k$ is the number of nodes in a complete graph
(note that link distances obey the triangle inequality and $k=n(M)$ in our problem).
To the best of our knowledge, the latest result is related to the works of Sebo et al. \cite {Sebo_2014}
and Hoogeveen \cite {Hoogeveen_1991}
for the path version, i.e., finding Hamiltonian Path without any Prespecified endpoints (HPP).

More precisely, it is not necessary to insert exactly two line segments in the last step of the original algorithm,
i.e., we modify this step to insert at most two line segments.
With this modification, $|\Pi'|$ is reduced proportional to $n(M)$, where $\Pi'$ is
an approximate minimum link visibility path.
Also, the minor changes are applied to the other steps.
The following modified algorithm computes $\Pi'$ for $P$
(functions $OneLink$, $TwoLink$, $Link$ and $Join$ are defined bellow after the algorithm):
\begin{enumerate}
\item [{$1)$}] Sort the vertices of $P$ in clockwise order.
               
\item [{$2)$}] Compute the set of nonredundant cuts $\mathcal{C}$.
               
\item [{$3)$}] If all cuts in $\mathcal{C}$ intersect, return one of these cuts.
               
\item [{$4)$}] Sort the endpoints in $\mathcal{C}$, in counterclockwise order.
               
\item [{$5)$}] Compute a maximum independent set of $\mathcal{C}$, $M=\{c_1,c_2,...,c_k\}$
               and its associated set of special regions $R=\{R_1,R_2,...,R_k\}$.
               
\item [{$6)$}] Compute $SPM(c_1),SPM(c_2),...,SPM(c_k)$ with the value of each face
               and prepare these maps for \emph{point location} queries.
               Find the link distance between each pair of cuts in $M$.
               
\item [{$7)$}] Construct the complete graph $G$ whose node set is $M$, and to each edge of it assign the link distance
               between the corresponding cuts.
               
\item [{$8)$}] Apply a well-known approximation algorithm
               (such as Sebo et al. or Hoogeveen, see \cite {Sebo_2014} or \cite {Hoogeveen_1991})
               for finding HPP with respect to $G$.
               Let $H$ be such a Hamiltonian path as the output of this algorithm.
               
\item [{$9)$}] Let $c_h,c_i,...,c_j \in M$ ($1 \leq h,i,..,j \leq k$) be the corresponding sequence of cuts in $H$.
               Rename the indices of cuts in this sequence as $c_1,c_2,...,c_k \in M$.
               Also, update the sequences in the set $R$ and the set of SPMs constructed in Step 6.
               
\item [{$10)$}] 
                Let $A_1=f_{1,2} \cap R_1$ and $\Pi'=Null$.
                
                If $A_1=\emptyset$, $\{\Pi'=OneLink(R_1,f_{1,2}); A_1=f_{1,2}\}$.
                
                For $i=1$ to $k-2$ do as follows:
                \begin{enumerate}
                \item [{$1)$}] Compute $Cellmin_{i+1,i,i+2}$.
                
                \item [{$2)$}] If $n(Cellmin_{i+1,i,i+2})=1$
                (cases $C(a)$ or $C(b)$ or $C(c_{3_1})$ or $C(c_{3_2})$)
                
                $\{$
                
                $C_l \in Cellmin_{i+1,i,i+2}$ ($C_l=f_{i+1,i} \cap f_{i+1,i+2}$);
                
                $A_{i+1}=C_l \cap R_{i+1}$;
                
                if $A_{i+1} \neq \emptyset$, $Join(\Pi',A_i,A_{i+1})$, otherwise,
                $\{$
                $T_1=f_{i+1,i} \cap R_{i+1}; T_2=f_{i+1,i+2} \cap R_{i+1}$;
                
                if $T_1=T_2=\emptyset$ (cases $C(c_{3_1})$ or $C(c_{3_2})$)
                
                $\{Join(\Pi',A_i,f_{i+1,i},TwoLink(f_{i+1,i},R_{i+1},f_{i+1,i+2})); A_{i+1}=f_{i+1,i+2}\}$;
                
                if ($T_1 \neq \emptyset$ and $T_2 \neq \emptyset$), (case $C(c_{3_1})$)
                $\{Join(\Pi',A_i,T_1,OneLink(T_1,T_2)); A_{i+1}=T_2\}$;
                
                if ($T_1=\emptyset$ and $T_2 \neq \emptyset$),
                $\{Join(\Pi',A_i,F_{i+1,i},Link(F_{i+1,i},T_2)); A_{i+1}=T_2\}$;
                
                if ($T_1 \neq \emptyset$ and $T_2=\emptyset$),
                $\{Join(\Pi',A_i,T_1,Link(T_1,F_{i+1,i+2})); A_{i+1}=F_{i+1,i+2}\}$
                $\}$
                
                $\}$.
                
                \item [{$3)$}] If $n(Cellmin_{i+1,i,i+2})=2$ (cases $C(c_1)$ or $C(c_2)$)
                
                $\{$
                
                $C_l,C_l' \in Cellmin_{i+1,i,i+2}$ ($C_l \subseteq f_{i+1,i}, C'_l \subseteq f_{i+1,i+2}$);
                $A_{i+1}=C_l \cap R_{i+1}; A'_{i+1}=C_l' \cap R_{i+1}$;
                
                if $A_{i+1}=A'_{i+1}=\emptyset$,
                $\{Join(\Pi',A_i,f_{i+1,i},TwoLink(f_{i+1,i},R_{i+1},f_{i+1,i+2})); A_{i+1}=f_{i+1,i+2}\}$;
                
                if $A'_{i+1} \neq \emptyset$,
                $\{Join(\Pi',A_i,F_{i+1,i},OneLink(F_{i+1,i},A'_{i+1})); A_{i+1}=A'_{i+1}\}$,
                
                otherwise, if $A_{i+1} \neq \emptyset$,
                $\{Join(\Pi',A_i,A_{i+1},OneLink(A_{i+1},F_{i+1,i+2})); A_{i+1}=F_{i+1,i+2}\}$
                
                $\}$.
                \end{enumerate}
                
                Let $A_k=f_{k,k-1} \cap R_k$.
                
                If $A_k=\emptyset$, $Join(\Pi',A_{k-1},f_{k,k-1},OneLink(f_{k,k-1},R_k))$,
                otherwise, $Join(\Pi',A_{k-1},A_k)$.
                
                Return $\Pi'$.
\end{enumerate}

Note that $OneLink(X,Y)$ provides a one link path from $X$ to $Y$ and returns this link, also,
$TwoLink(X,Y,Z)$ provides a two links path from $X$ to $Y$, then $Y$ to $Z$ and returns this path
($X$, $Y$ and $Z$ are simple polygons such that $X \cap Y=\emptyset$ and $Y \cap Z=\emptyset$).
$Link(X,Y)$ provides a minimum link path from $X$ to $Y$ with link distance one or two and returns this path.
Let $\pi_x$ and $\pi_y$ be two polygonal paths ($|\pi_y|=$ 1 or 2), where the endpoint of $\pi_x$ (called $x$)
lies in $X$ and the starting point of $\pi_y$ (called $y$) lies in $Y$ (if $\pi_x=Null$, then $x$ will be an arbitrary
point from $X$).
$Join(\pi_x,X,Y,\pi_y)$, finds a minimum link path between $x$ and $y$ (called $\pi_{x,y}$) and
sets $\pi_x=\pi_x+\pi_{x,y}+\pi_y$ ($'+'$ joins paths).
Similarly, $Join(\pi_x,X,Y)$ is defined, i.e., let $\pi_y=Null$ and $y$ be an arbitrary point from $Y$.

To implement the function $Link$, we use the \emph{geodesic hourglass}
between convex polygons presented in \cite {Arkin_1995}.
If the hourglass is \emph{open}, a path with link distance one is returned, otherwise, using the common \emph{apex} of
the two created \emph{funnels}, a path with link distance two is returned (see \cite {Guibas_1989}).
Since the arguments of this function are simple polygons,
we must use their convex chain as follows.
Remember that each $R_i$ is a simple polygon and $\partial R_i$ consists of two parts:
one on $\partial Pocket(c_i)$ (can be concave) and the other inside $Pocket(c_i)$,
which is a convex chain ($c_i \in M$).
We call $R_i$ \emph{partial convex}.
It is easy to see that $f_{i,j} \cap R_i$ is also partial convex ($1 \leq j \leq k$).
Thus, $Link(F_{i+1,i},T_2)$ and $Link(T_1,F_{i+1,i+2})$ can be computed in $O(\log n_i \log n)$ time \cite {Arkin_1995}.
This can be done by computing a minimum link path between the closest window of $F_{i+1,i}$(or $F_{i+1,i+2}$), which intersects
$c_{i+1}$, and the convex chain of $T_2$(or $T_1$).
Note that since $c_{i+1}$ crosses both $F_{i+1,i}$ and $F_{i+1,i+2}$, the link distance of this path is at most two.

Similarly, the functions $OneLink$ and $TwoLink$ are implemented ($A_i$ and $A'_i$ are convex or partial convex).
More precisely, $OneLink(R_1,f_{1,2})$ and $OneLink(f_{k,k-1},R_k)$ can connect $R_1$ to $f_{1,2}$
and $f_{k,k-1}$ to $R_k$ by $c_1$ and $c_k$, respectively.
$OneLink(T_1,T_2)$ can connect $T_1$ and $T_2$ inside the convex chain of $R_{i+1}$ by one link in the case $C(c_{3_1})$.
$OneLink(F_{i+1,i},A'_{i+1})$ and $OneLink(A_{i+1},F_{i+1,i+2})$ can connect their arguments by one link
due to the construction of $C'_l$ and $C_l$, respectively.
Finally, $TwoLink(f_{i+1,i},R_{i+1},f_{i+1,i+2})$ can connect $f_{i+1,i}$ to $R_{i+1}$ by $c_{i+1}$ and then
$R_{i+1}$ to $f_{i+1,i+2}$, again by $c_{i+1}$.
This shows a path with link distance one and two always exists,
where we use the functions $OneLink$ and $TwoLink$ in the algorithm, respectively.

To implement the function $Join$ (both versions), $\pi_{x,y}$ must be computed.
According to Step 6 and \cite {Suri_1990}, this can be done in $O(\log n+|\pi_{x,y}|)$ time.
Since $0 \leq |\pi_y| \leq 2$, the total time complexity of this function would be $O(k \log n + |\Pi'|)$ for all iterations.

\begin{figure}
	\centering
	\includegraphics[width=15cm,keepaspectratio=true]{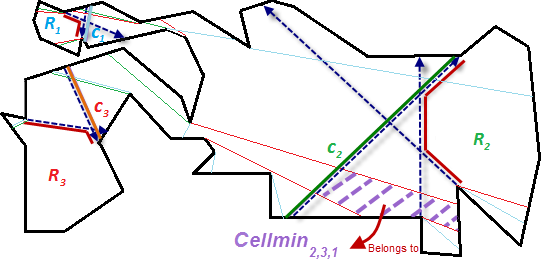}
	\caption{$SPM(c_1), SPM(c_2)$ and $SPM(c_3)$ create \emph{Cells} for $c_1, c_2$ and $c_3 \in M$}
	\label{fig:Regions}
\end{figure}

Now, let us briefly describe the algorithm by walking through the example depicted in Figure~\ref{fig:Regions}.
In Steps 1 to 5, the sets $M=\{c_1,c_2,c_3\}$ and $R=\{R_1,R_2,R_3\}$ are computed.
The complete graph $G$ with the vertices $\{c_1,c_2,c_3\}$ and the edges
$\{\overline{c_1c_2},\overline{c_2c_3},\overline{c_3c_1}\}$ with labels $\{2,1,3\}$, respectively,
is constructed in Steps 6 and 7.
Suppose that the sequence $c_3,c_2,c_1$ is produced after Step 8 as the input of Step 9.
Again, in Step 9, this sequence is renamed (in this example, we will be working with the original sequence).
Finally, in Step 10, $R_3$ is connected to $f_{3,2}$ by one link.
Then, $Cellmin_{2,3,1}$ is computed. Since $n(Cellmin_{2,3,1})=1$ and $C_l \cap R_2=\emptyset$ and
$T_1=f_{2,3} \cap R_2 = \emptyset$ and $T_2=f_{2,1} \cap R_2 \neq \emptyset$,
$F_{2,3}$ is connected to $T_2$ by one link.
Then, $f_{3,2}$ is connected to $F_{2,3}$ by the $Join$ function.
As $A_3=f_{1,2} \cap R_1 \neq \emptyset$, $T_2$ is connected to $A_3$ by the $Join$ function.
So, $\Pi'$ would be the constructed path from $R_3$ to $R_1$.

It is easy to modify the above algorithm to compute a tour
(a path whose start and end points coincide) instead of a path.
Let $\gamma$ be the approximation factor of an $O(k^3)$ time approximation algorithm
for finding HPP (or HT)\footnote
{The best known value of $\gamma$ is $1.5$ for finding HPP \cite {Hoogeveen_1991,Sebo_2014}
and $1.4$ for finding HT (Hamiltonian Tour) \cite {Sebo_2014}.} with respect to $G$,
where $k$ is the number of nodes in $G$, i.e., $k=n(M)$.
Therefore, for a minimum link visibility path (tour) $\Pi$:
$\sum_{i=1}^{m} |\pi_{x,y}| < \gamma |\Pi|$, and as discussed in the previous section
$m \leq |\Pi|$,
where $m=k-1$, if $\Pi$ is a path and $m=k$, if $\Pi$ is a tour ($|\Pi|=Opt$).
Let $a_l=(\sum_{i=1}^{m-1} |\pi_{y}|+$(the number of links added to connect $R_1$ to $f_{1,2}$ and $f_{k,k-1}$ to $R_k$))
in the path version and $a_l=(\sum_{i=1}^{m} |\pi_{y}|)$ in the tour version.
Since for each cut in $M$ at most two links have been added by the functions $OneLink$, $TwoLink$ and $Link$,
$a_l/m$ is a positive rational number no more than 2 for both versions and we have:

\begin{center}
$|\Pi'|=\sum_{i=1}^{m} |\pi_{x,y}| + a_l < \gamma |\Pi| + (a_l/m)m \leq (\gamma + a_l/m)|\Pi|=(\gamma + a_l/m)Opt$
\end{center}

The time complexity of the algorithm is computed as follows.
Steps 1 and 4 take $O(n \log n)$.
As in the previous arguments, Steps 2, 3 and 5 take altogether $O(n^2)$.
Computation and preprocessing of SPMs take $O(kn)$ in Step 6 (\cite {Suri_1990, Edelsbrunner_1986}).
Also, the link distances can be computed in $O(k^2 \log n)$ by answering the
\resizebox{0.0375\hsize}{!}{$\binom{k}{2}$}$=O(k^2)$
queries in this step \cite {Suri_1990}.
Step 7 takes $O(k^2)$.
According to \cite {Hoogeveen_1991, Sebo_2014}, Step 8 takes $O(k^3)$.
Step 9 takes $O(k)$.
Finally, we will show that Step 10 can be done in $O(n^2)$.
Hence, the overall time taken by the algorithm is $O(kn^2)$ ($k=O(n)$).

From the previous section, we know the number of vertices of $R_i$, $C_l$, $C'_l$ and $f_{i,j}$
($1 \leq i,j \leq k$) are $O(n_i)$, where $\sum_{i=1}^{k} n_i=O(n)$.
Thus, $f_{i,j} \cap R_i$, $C_l \cap R_i$ and $C'_l \cap R_i$ can be computed in $O(n_i)$ time
using the algorithm given in Chazelle \cite {Chazelle_1991}.
Computation of $Cellmin_{i+1,i,i+2}$ and its corresponding faces take $O(n)$ time ($1 \leq i \leq k-2$).
The total time complexity of the functions $OneLink$, $TwoLink$ and $Link$ for all iterations is
$\sum_{i=1}^{k} O(\log n_i \log n)=O(\log^2 n)$
(also, $O(n)$ time preprocessing of $P$ for shortest path queries is needed \cite {Arkin_1995}).
Similarly, since $O(|\Pi'|)=O(|\Pi|)=O(n)$, the time complexity of the function $Join$ is $O(n \log n)$
for all iterations.
This proves the following theorem:

\begin{theorem}
Let $P$ be a watchman polygon with $n$ vertices and $k$ the size of a maximum independent set of cuts.
It is always possible to construct polygonal visibility paths (or tours) of link-length at most
$(\gamma+a_l/(k-1))Opt$ (or $(\gamma+a_l/k)Opt$) in time $O(kn^2)$,
where $Opt$ is the link-length of an optimal solution for path (or tour),
$a_l$ is the number of added links in the algorithm and $\gamma$ is the approximation factor of an $O(k^3)$ time approximation algorithm for finding HPP (or HT).
\end{theorem}

\section*{Acknowledgements}
We sincerely thank
Dr Hadi Shakibian and Dr Ali Rajaei
for their kind help and valuable comments.


\begin{thebibliography}{99}

\bibitem{Alsuwaiyel_1993}
M. H. Alsuwaiyel, D. T. Lee, Minimal link visibility paths inside a simple polygon,
\textit{Computational Geometry Theory and Applications},
\textbf{3}(1), (1993), 1-25.

\bibitem{Alsuwaiyel_1995}
M. H. Alsuwaiyel, D. T. Lee, Finding an approximate minimum-link visibility path inside a simple polygon,
\textit{Information Processing Letters},
\textbf{55}(2), (1995), 75-79.

\bibitem{Arkin_2003}
E. M. Arkin, J. S. B. Mitchell, C. D. Piatko, Minimum-link watchman tour,
\textit{Information Processing Letters},
\textbf{86}(4), (2003), 203-207.

\bibitem{Arkin_1995}
E. M. Arkin, J. S. B. Mitchell, S. Suri, Logarithmic-time link path queries in a simple polygon,
\textit{International Journal of Computational Geometry and Applications},
\textbf{5}(4), (1995), 369-395.

\bibitem{Chazelle_1991}
B. Chazelle, Triangulating a simple polygon in linear time,
\textit{Discrete and Computational Geometry},
\textbf{6}(3), (1991), 485-524.

\bibitem{Christofides_1976}
N. Christofides, Worst case analysis of a new heuristic for the traveling salesman problem,
\textit{Technical Report, Graduate School of Industrial Administration},
\textbf{CMU}(388), (1976).

\bibitem{Edelsbrunner_1986}
H. Edelsbrunner, L. J. Guibas, J. Stolfi, Optimal point location in a monotone subdivision,
\textit{SIAM Journal on Computing},
\textbf{15}(2), (1986), 317-340.

\bibitem{Guibas_1989}
L. J. Guibas, J. Hershberger, Optimal shortest path queries in a simple polygon,
\textit{Journal of Computer and System Sciences},
\textbf{39}(2), (1989), 126-152.

\bibitem{Hoogeveen_1991}
J. A. Hoogeveen, Analysis of Christofides’ heuristic: Some paths are more difficult than cycles,
\textit{Operations Research Letters},
\textbf{10}(5), (1991), 291-295.

\bibitem{Sebo_2014}
A. Sebo, J. Vygen, Shorter Tours by Nicer Ears: 7/5-approximation for graphic TSP, 3/2 for the path version, and 4/3 for two-edge-connected subgraphs,
\textit{Combinatorica},
\textbf{34}(5), (2014), 597-629.

\bibitem{Suri_1990}
S. Suri, On some link distance problems in a simple polygon,
\textit{IEEE transactions on Robotics and Automation},
\textbf{6}(1), (1990), 108-113.

\bibitem{Tan_2007}
X. Tan, A linear-time 2-approximation algorithm for the watchman route problem for simple polygons,
\textit{Theoretical Computer Science},
\textbf{384}(1), (2007), 92-103.

\bibitem{Zarrabi_2020_1}
M. R. Zarrabi, N. M. Charkari, Query-points visibility constraint minimum link paths in simple polygons,
\textit{arXiv preprint arXiv:2004.02220}, (2020).

\bibitem{Zarrabi_2020_2}
M. R. Zarrabi, N. M. Charkari, A simple proof for visibility paths in simple polygons,
\textit{arXiv preprint arXiv:2004.02227}, (2020).

\end{thebibliography}
\end{document}